\documentclass[pre,aps,twocolumn,superscriptaddress,floatfix,amsmath,amssymb,showpacs]{revtex4}

\usepackage{graphicx}
\usepackage{bm}

\preprint{LA-UR 09-01331, LA-UR 09-03784}

\begin{document}

\title{
  An \emph{ab initio} method for locating characteristic potential energy
  minima
}

\date{\today}

\author{E. Holmstr\"{o}m}
\email{eholmstrom@uach.cl}
\affiliation{Instituto de F\'{i}sica, Universidad Austral de Chile, Valdivia,
Chile}

\author{N. Bock}
\author{Travis~B. Peery}
\affiliation{Theoretical Division, Los Alamos National Laboratory, Los Alamos,
NM 87545, USA}

\author{R. Liz\'{a}rraga}
\affiliation{Instituto de F\'{i}sica, Universidad Austral de Chile, Valdivia,
Chile}

\author{G. De~Lorenzi-Venneri}
\author{Eric~D. Chisolm}
\author{Duane~C. Wallace}
\affiliation{Theoretical Division, Los Alamos National Laboratory, Los Alamos,
NM 87545, USA}

\begin{abstract}

  It is possible in principle to probe the many--atom potential surface using
  density functional theory (DFT). This will allow us to apply DFT to the
  Hamiltonian formulation of atomic motion in monatomic liquids
  [\textit{Phys.~Rev.~E} {\bf 56}, 4179 (1997)]. For a monatomic system,
  analysis of the potential surface is facilitated by the random and symmetric
  classification of potential energy valleys. Since the random valleys are
  numerically dominant and uniform in their macroscopic potential properties,
  only a few quenches are necessary to establish these properties.  Here we
  describe an efficient technique for doing this. Quenches are done from
  easily generated ``stochastic'' configurations, in which the nuclei are
  distributed uniformly within a constraint limiting the closeness of
  approach. For metallic Na with atomic pair potential interactions, it is
  shown that quenches from stochastic configurations and quenches from
  equilibrium liquid Molecular Dynamics (MD) configurations produce
  statistically identical distributions of the structural potential energy.
  Again for metallic Na, it is shown that DFT quenches from stochastic
  configurations provide the parameters which calibrate the Hamiltonian. A
  statistical mechanical analysis shows how the underlying potential
  properties can be extracted from the distributions found in quenches from
  stochastic configurations.

\end{abstract}

\pacs{05.70.Ce, 61.20.Gy, 71.15.Nc}
\keywords{}

\maketitle

\section{Introduction}
\label{sec:Introduction}

The potential energy surface underlying the motion of a monatomic
liquid is composed of intersecting valleys in the many--atom
configuration space. In Vibration--Transit (V-T) theory, the atomic
motion is described by vibrations within a single valley, interspersed
by transits, which carry the system across intervalley intersections
\cite{PhysRevE.56.4179, 0953-8984-13-37-201}. The pure vibrational
motion is described by a tractable Hamiltonian which accounts for the
dominant part of the liquid's thermodynamic properties. The remaining
transit contribution is complicated but small. The vibrational
Hamiltonian is calibrated by potential parameters evaluated at local
minima (called {\emph{structures}) of the potential surface.  These
have been previously evaluated for a model of liquid Na based on an
interatomic pair potential \cite{lorenzi-venneri:041203}. Our goal now
is to introduce first principles electronic structure calculations
within density functional theory (DFT) to calculate the structural and
vibrational parameters of the V-T Hamiltonian.

In recent years, DFT has been used successfully in liquid studies to support
the development of \emph{ab initio} Molecular Dynamics (MD). The original
Car--Parrinello method \cite{PhysRevLett.55.2471} was successfully applied to
the melting of Si \cite{PhysRevLett.74.1823}. Calculations of the MD
trajectory on the adiabatic (electronic ground state) potential surface have
been performed for a broad spectrum of elemental liquids
\cite{PhysRevB.47.558, PhysRevB.48.13115, JNonCrystallineSolids.312_314.52}.
This work has provided accurate and physically revealing results for Al
\cite{PhysRevB.57.8223}, Fe \cite{PhysRevB.61.132}, and Ge
\cite{PhysRevB.67.104205}. The techniques we employ to calculate the supercell
ground state energy and Hellmann--Feynman forces are quite similar to those of
Kresse et al.~\cite{ComputationalMaterialScience.6.15}. On the other hand,
rather than proceed to MD calculations, our objective is to calibrate the
(dominant) vibrational part of the liquid dynamics Hamiltonian from properties
of local potential minima.  We believe that the Hamiltonian formulation will
usefully complement \emph{ab initio} MD in the study of dynamical properties
of liquids.

The procedure of quenching a system to its potential energy minima was
introduced by Stillinger and Weber \cite{PhysRevA.25.978, weber:2742,
FrankH.Stillinger09071984}, and has become a valuable technique for
studying systems with interatomic potentials. The traditional method
is to quench to many structures from an equilibrium MD trajectory, and then
use statistical mechanics to extract the underlying statistical properties of
the potential surface \cite{0953-8984-12-29-325, PhysRevLett.83.3214}. In V-T
theory, however, we need very few structures---in principle only one for a
given liquid at a given volume.  Hence the traditional procedure for finding
structures is inefficient, as a large number of MD iterations is required to
bring the system initially to equilibrium as well as to avoid unwanted
correlations between quenches.  The problem is especially severe for DFT,
where each iteration requires a converged total energy calculation, which is
computationally very costly.

We propose a simpler and more efficient method to probe the underlying
potential landscape.  Rather than quench from equilibrium MD
configurations, we quench from configurations that are independent of
interatomic interactions, and very fast to generate. Our purpose here
is to demonstrate two properties of this quench method: that it is
capable of producing the entire distribution of potential energy
minima, and when used with DFT it can achieve our goal of \textit{ab
initio} calibration of the Hamiltonian.

To perform the calibration, we will interpret structural data via the
original hypothesis of V-T theory \cite{PhysRevE.56.4179}: The
potential energy valleys are classified as random and symmetric. The
random valleys numerically dominate the liquid statistical mechanics
as $N \rightarrow \infty$, and they all have the same potential energy
properties as $N \rightarrow \infty$; hence any one such valley may be
used to calibrate the Hamiltonian. The symmetric valleys have a broad
range of potential energy properties, but make an insignificant
contribution to the liquid statistical mechanics as $N \rightarrow
\infty$. This hypothesis has been verified for the pair potential
model of liquid Na at $N = 500$ \cite{lorenzi-venneri:041203}, and
work in progress extends this verification to $N = 4000$
\cite{N-Dependent-Quenches}.

The quench calculations are carried out for metallic Na at the density of the
liquid at melt, using our model interatomic potential, and also using DFT. The
Na potential was derived in pseudopotential perturbation theory, with an added
Born--Mayer repulsion, and was calibrated from experimental crystal data at
zero temperature and pressure \cite{PhysRev.176.832}. The potential has since
been shown to give excellent results for a broad range of experimental
properties of crystal and liquid Na (for a partial summary, see
\cite{Wallace:SPCL}). While there is no doubt DFT will provide accurate total
energy results, we shall still need to verify that DFT quenches arrive at
random structures, not symmetric ones. This verification will be accomplished
with the aid of the Na interatomic potential results.

Our application of DFT to liquid dynamics theory is being pursued for a number
of nearly--free--electron metals and transition metals. A preliminary report
has been presented on results for Na and Cu \cite{bock_march_meeting_2008}. We
have not attempted to study elemental liquids whose equilibrium configurations
are influenced by molecular bonding. Examples are As, Se, and Te, which have
strong and weak bonds, and Ge, whose liquid structure shows a contribution
from covalent crystal bonds. This anisotropic bonding will complicate the
random structures underlying the motion in such liquids.  This complication
remains beyond the scope of the present work.  Moreover, since we have not yet
presented structural data for other liquid metals, the present conclusions are
strictly valid only for liquid Na.  The results are expected to apply to many
elemental liquids, perhaps all of them, but this extension is not demonstrated
here.

We consider a system of atoms in a cubic box with periodic boundary
conditions.  We construct configurations in which the nuclei are
distributed uniformly over the box, within a constraint limiting the
closeness of approach of any pair.  These are called \emph{constrained
stochastic}, or simply \emph{stochastic}, configurations.  The
procedure is described in Sec.~\ref{sec:generating}, and the spatial
uniformity is verified by means of pair distribution functions. In
Sec.~\ref{sec:properties}, our twofold purpose is addressed by two
separate quench studies. In Sec.~\ref{sec:validatestochastic}, the Na
pair potential is used to quench both equilibrium MD configurations
and stochastic configurations.  Comparison of the results will validate
the use of stochastic configurations. In Sec.~\ref{sec:calibrateH},
DFT is used to quench Na from stochastic configurations. Comparison of
the results with pair potential results will confirm that the DFT
structures are random and therefore calibrate the Hamiltonian.  In
Sec.~\ref{sec:extracting}, relations are derived between quenched
distributions and the densities of states in the underlying potential
energy surface. This analysis provides the statistical mechanical
framework for interpreting results of the present quench
technique. Our conclusions are summarized in
Sec.~\ref{sec:conclusions}.

\section{Generating stochastic configurations}
\label{sec:generating}

Only minimal information is required to generate stochastic
configurations: the number of atoms $N$ and the system volume $V$,
plus a distance of closest approach which is described below. Nothing
of boundary conditions or the system potential has to be specified;
however, after the stochastic configuration is constructed, for all
further calculations periodic boundary conditions are used.

\begin{figure}
\includegraphics[width=0.9\linewidth]{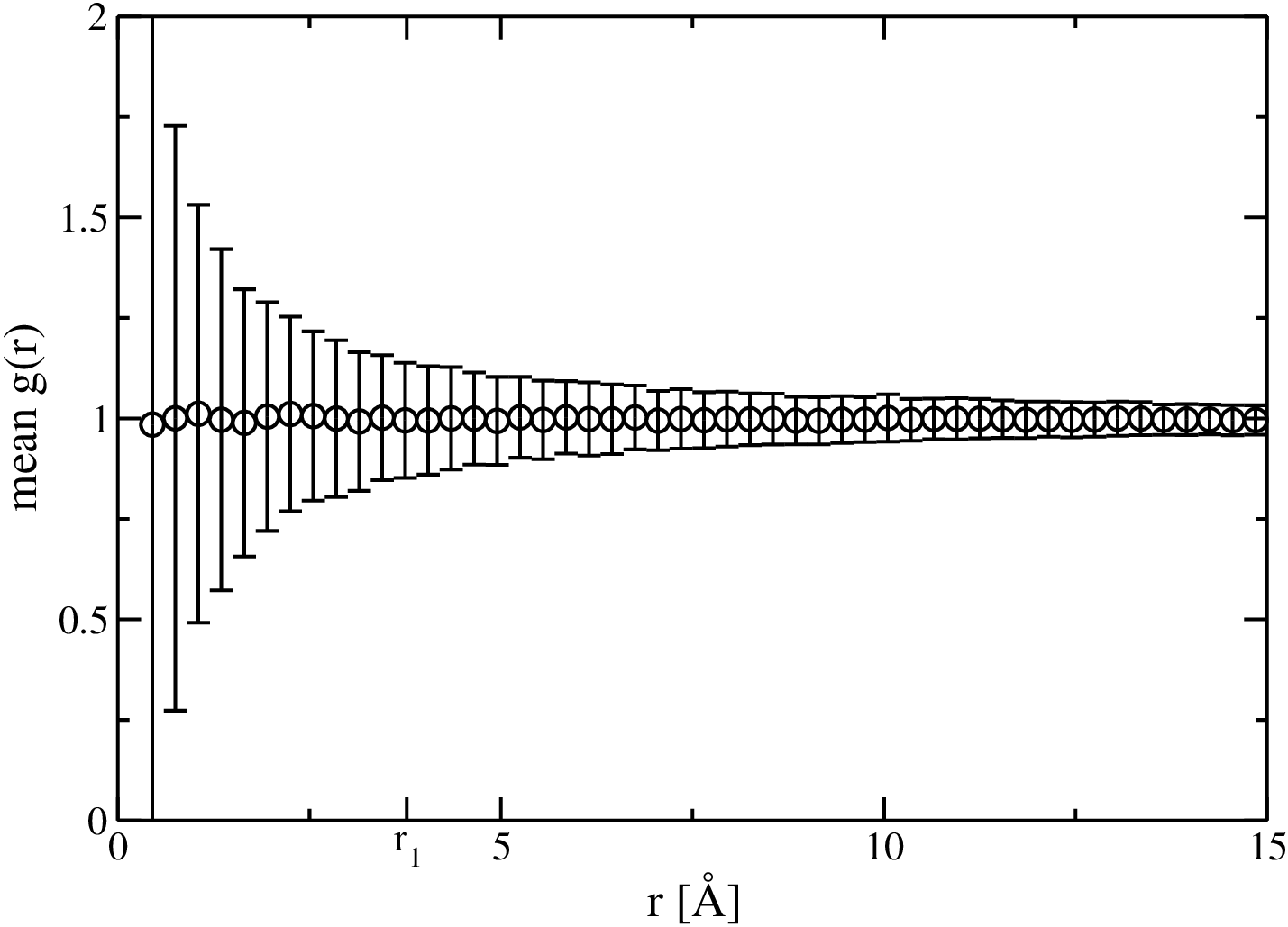}
\caption{
  \label{fig:gOfR_constant_binsize}
  Mean and standard deviation of the pair distribution functions $g(r)$ for
  1000 stochastic configurations of $N = 500$ atoms. The error bars
  indicate the standard deviation per histogram bin. The center of the nearest
  neighbor peak in $g(r)$ for the quenched configuration of Na, shown in
  Fig.~\ref{fig:gOfR}, is at $r_{1} = 3.77$ \AA.}
\end{figure}

For a cubic cell with volume $V$, we construct a configuration by
choosing the particle coordinates at random over the cell. Randomness
is important, as we shall use it in Sec.~\ref{sec:extracting} to
determine the statistical weight factors for stochastic
configurations. Next, a configuration is discarded if any two atoms
are closer than a distance $d$. This is done for practical reasons:
The self consistent field (SCF) calculation of DFT will not converge
if atoms are too close to each other, and the pair potential at very
small radii could lead to numerical instabilities in the conjugate
gradient method due to the repulsive core. In practice, the excluded
space can be very small.  For Na we choose $d = 0.4$ \AA, so the
relative excluded space $\left(4 \pi / 3 \right) d^{3}/V_{A}$, where
$V_{A}$ is the volume per atom, is only $6.5 \times 10^{-3}$. Hence
the stochastic configurations are expected to be spatially uniform to
a very good approximation.

To test the uniformity of stochastic configurations, we examine their
pair distribution functions $g(r)$. The conditional probability
density $g(r)$ is constructed as follows: Pick a system atom as
central atom and denote its position by $r = 0$. Make a set of bins
labeled $b = 1, 2, \cdots$, in the form of concentric shells. Bin $b$
has inner radius $r_{b}$, outer radius $r_{b+1}$, and volume $V_{b} =
(4 \pi / 3) \left( r_{b+1}^{3} - r_{b}^{3} \right)$. The pair
distribution function $g(r)$ has histogram $n_{b} \left( V_{A}/ V_{b}
\right)$, where $n_{b}$ is the number of atoms in bin $b$. Given the
small size of $d$ described in the previous paragraph, we expect
$g(r)$ to be nonzero even at relatively small radii. It is therefore
important to normalize the bin count of bin $b$ with the correct
volume of the bin, instead of using the approximate volume $4 \pi
r_{b}^{2} \Delta r_{b}$ as is often done. The bin contents are then
averaged over the choice of each system atom as the central
atom. While the bin radii are arbitrary, we usually take either
$\Delta r_{b} = r_{b+1}-r_{b} =$ constant or $V_{b} =$ constant.

\begin{figure}
\includegraphics[width=0.9\linewidth]{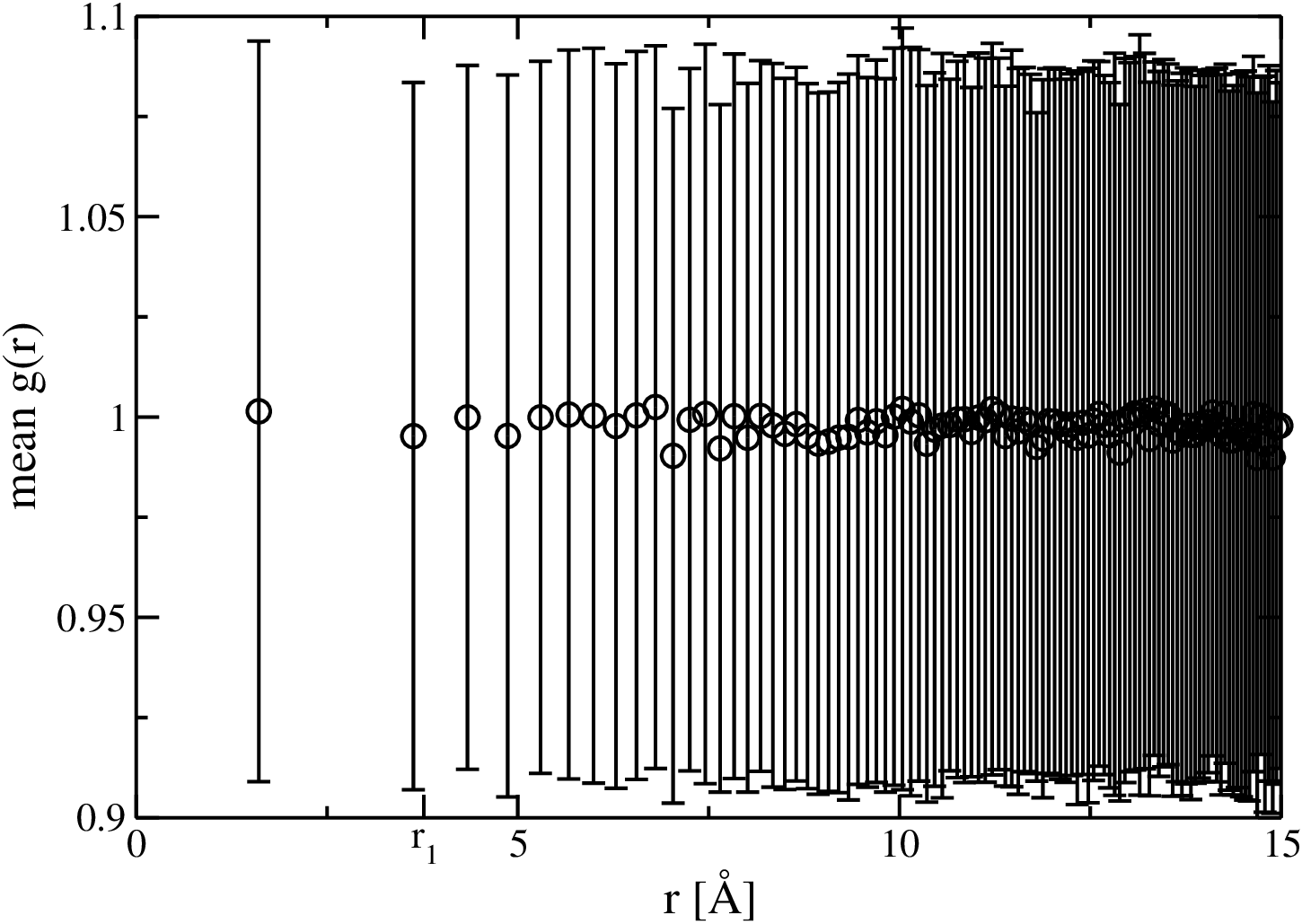}
\caption{
  \label{fig:gOfR_constant_volume}
  Mean and standard deviation of the pair distribution functions $g(r)$ for
  1000 stochastic configurations of $N = 500$ atoms. The bins have a
  constant volume of $V_{b} = (1/8) V_{A}$, where $V_{A}$ is the atomic
  volume of Na. The center of the nearest neighbor peak in $g(r)$ for the quenched
  configuration, shown in Fig.~\ref{fig:gOfR}, is at $r_{1} = 3.77$
  \AA. For this particular binning volume, we have only 2 bins below the
  nearest neighbor peak.}
\end{figure}

For Na at $N = 500$, we constructed 1000 stochastic configurations and
the $g(r)$ histogram for each. With $\Delta r_{b} =$ constant, the
mean and the standard deviation of the $g(r)$ histogram are shown in
Fig.~\ref{fig:gOfR_constant_binsize}.  The blank space at small $r$ is
the empty sphere of radius $d$. The scatter at small $r$ reflects the
decreasing $V_{b}$ as $r$ decreases, and the corresponding decrease in
$n_{b}$. With $V_{b} =$ constant, the mean and standard deviation of
the $g(r)$ histogram is shown in Fig.~\ref{fig:gOfR_constant_volume}.
There the distance between points increases as $r$ decreases, but the
standard deviation remains nearly constant because $n_{b}$ remains
nearly constant. The figures show the uniformity of stochastic
configurations for $r > d$, with $d$ being very small compared to the
mean nearest neighbor distance.

We also show the distribution of potential energies of the 1000
initial stochastic configurations. The mean of the distribution shown
in Fig.~\ref{fig:initial_energy_stochastic} is at 3.75 eV/atom, which
is considerably higher than the mean potential energy after the quench
(see Eq.~(\ref{eq:mean_energy})).

\section{Properties of the quenched structures}
\label{sec:properties}

\subsection{Validation of quenches from stochastic configurations}
\label{sec:validatestochastic}

Figs.~\ref{fig:distributionMD} and \ref{fig:distributionStochastic}
compare two distributions of $\Phi_{0}/N$, each from 1000 pair
potential quenches at $N = 500$.  Fig.~\ref{fig:distributionMD} is
obtained by steepest--descent quenches from equilibrium MD at 800
K. This figure is an extension of the work reported by
\cite{lorenzi-venneri:041203}.  Fig.~\ref{fig:distributionStochastic}
is obtained by conjugate gradient quenches from stochastic
configurations. We have verified the equivalence of the two quench
techniques for our system; for a related verification, see
\cite{chakravarty:206101, chakraborty:014507}.

\begin{figure}
\includegraphics[width=0.9\linewidth]{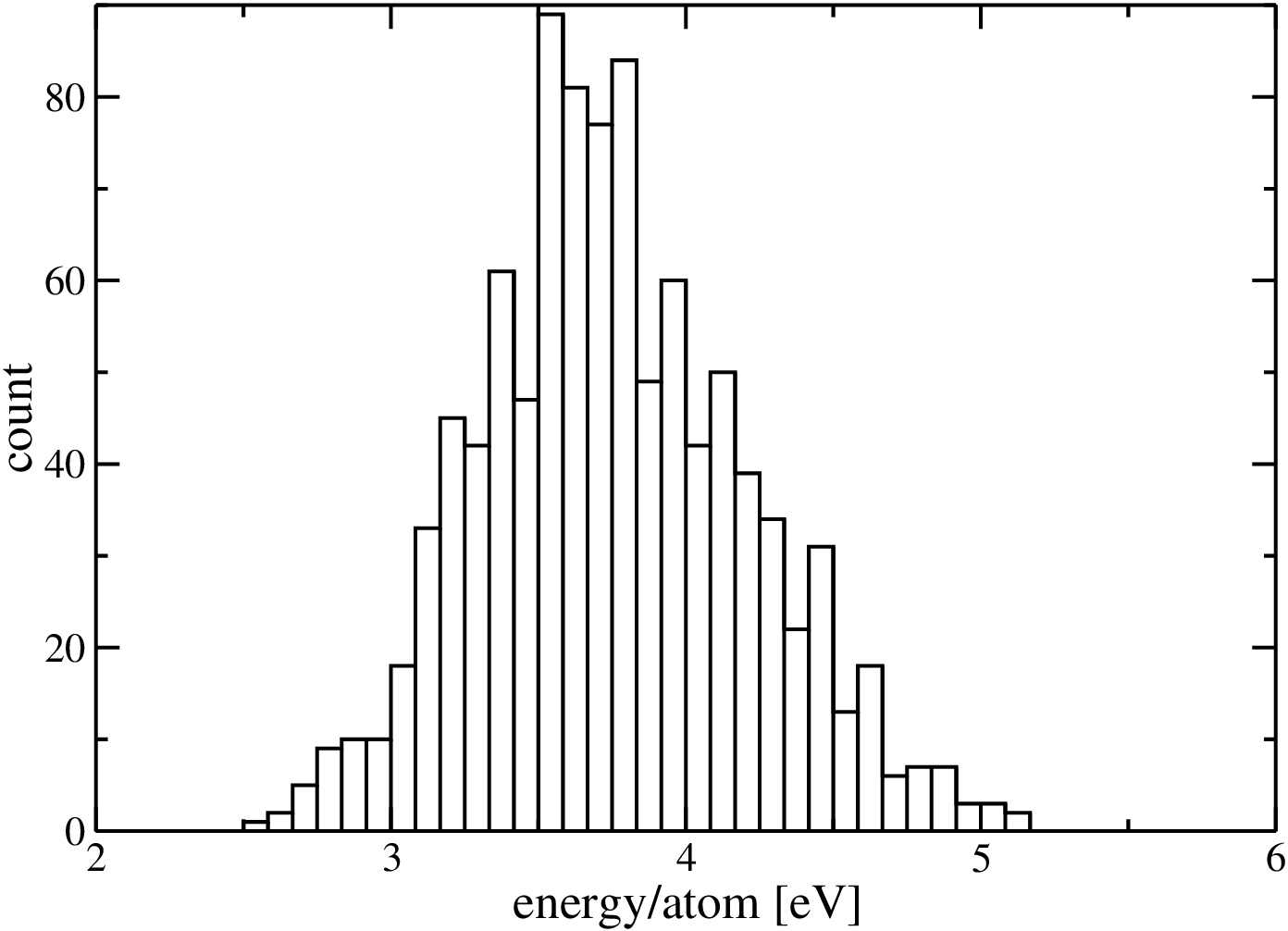}
\caption{
  \label{fig:initial_energy_stochastic}
  Potential energy distribution of 1000 stochastic configurations of $N = 500$
  Na atom systems.}
\end{figure}

In each histogram we see the distinct random and symmetric
distributions, consistent with the V-T hypothesis. The random
distribution is taken to be the dominant peak, out to where the
histogram vanishes on either side.  The symmetric structures are
interpreted as the isolated parts of the histogram that are located
outside the main peak on the low--energy side.

The random distribution is numerically dominant and very narrow. The
mean and standard deviation of each random distribution is given by
\begin{equation}
  \label{eq:mean_energy}
  \Phi_{0}/N = \left\{
    \begin{array}{l}
      -183.29 \pm 0.50 \mbox{ meV/atom (MD)} \\
      -183.37 \pm 0.50 \mbox{ meV/atom (stochastic)}.
    \end{array} \right.
\end{equation}
Note that the dominant volume--dependent part of the pair potential is
omitted here (see \cite{PhysRevE.59.2942}, Eq.~(1.1) and Fig.~1). For
this reason, most of the binding energy is missing from the pair
potential energies in Eq.~(\ref{eq:mean_energy}).  The mean value is
the most accurate approximation to the thermodynamic limit value that
we can make. The standard deviation is the error expected from
quenching only once and using that result. If $N$ is increased, the
mean value is expected to change slightly in converging to its
thermodynamic limit, while the standard deviation goes to zero as $N
\rightarrow \infty$.  For a physical measure of the difference in mean
values, we note that the main contribution to the liquid thermal
energy is the classical vibrational contribution $3 k_{B} T$, which is 95.90
meV/atom at $T_{m}$.  The difference in means is 0.08\% of this.
Experimental error in the thermal energy of elemental liquids at
$T_{m}$ is typically (0.1 - 0.5)\%.  Hence the two random
distributions in Figs.~\ref{fig:distributionMD} and
\ref{fig:distributionStochastic} are identical to better than
experimental error.

\begin{figure}
\includegraphics[width=0.9\linewidth]{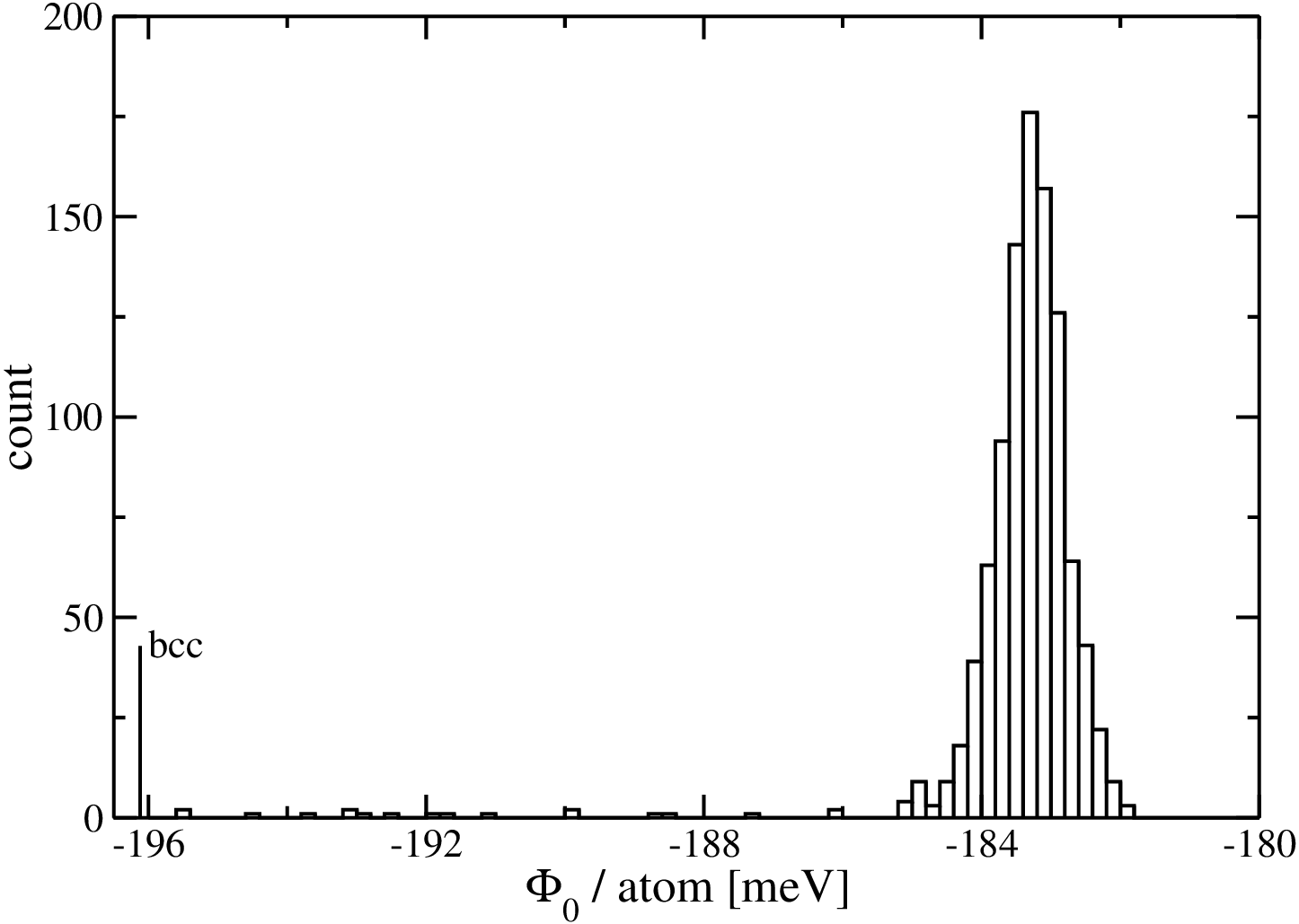}
\caption{
  \label{fig:distributionMD}
  Potential energy distribution of $N = 500$ Na atom systems after quenching
  from 1000 configurations drawn from a molecular dynamics trajectory at 800
  K, or $2.16 T_{m}$, using a steepest--descent quench method.}
\end{figure}

Performing so many quenches has allowed us for the first time to see a
clear and meaningful distribution of symmetric structures (compare for
example \cite{lorenzi-venneri:041203, PhysRevE.59.2955,
PhysRevE.59.2942}). Quenching from equilibrium MD yields 18 symmetric
structures out of 1000 quenches, Fig.~\ref{fig:distributionMD}.
Quenching from stochastic configurations yields 23 symmetric
structures in 1000 quenches, Fig.~\ref{fig:distributionStochastic}.
Very approximately, the symmetric distribution is constant and ranges
from the bcc crystal, $\Phi_{0}^{bcc}/N = -196.12$ meV/atom
\cite{PhysRevE.59.2942}, to the lower end of the random
distribution. This broad distribution with few structures is
consistent with the V-T hypothesis. If $N$ is increased the symmetric
distribution width is expected to remain the same, while the relative
number of symmetrics is expected to become negligible. In all these
properties, the symmetric distributions in
Figs.~\ref{fig:distributionMD} and \ref{fig:distributionStochastic}
are the same to statistical accuracy.

\subsection{Calibration of the V-T Hamiltonian}
\label{sec:calibrateH}

In order to calculate \textit{ab initio} the thermodynamic properties of
liquid Na for comparison with experiment, we quenched a stochastic
configuration at $V_{A} = 41.27$ \AA${}^{3}$ with DFT
\cite{bock_march_meeting_2008}. The normal mode frequency spectrum $g(\omega)$
was also calculated at this volume.  For almost all monatomic liquids, the
vibrational motion is nearly classical at $T \geq T_m$.  This means that the
essential information required from $g(\omega)$ is the logarithmic moment of
the frequencies, which provides the characteristic temperature $\theta_0$.
Hence for calculation of thermodynamic properties, the V-T Hamiltonian is
calibrated from $\Phi_0/N$ for the energy, and $\theta_0$ for the entropy.
Additional data which automatically accompanies the calculation of $g(\omega)$
will calibrate the Hamiltonian for nonequilibrium statistical mechanics.

The DFT calculation is done with the VASP code \cite{VASP}, using the
projector augmented wave (PAW) method in the generalized gradient
approximation (GGA) \cite{PhysRevB.50.17953, PhysRevB.59.1758}. The planewave
energy cutoff is 101.7 eV, the maximum core radius is 2.5 \AA, and the total
energy convergence criterion is $10^{-8}$ eV. We use a $\Gamma$--centered
Monkhorst--Pack grid \cite{PhysRevB.13.5188} with 14 $\bm{k}$--points in the
irreducible Brillouin zone. The total energy convergence for these parameters
was carefully verified. Note that it is the large size of the real--space
supercell which allows us to use few $\bm{k}$--points in comparison to the
large number (several thousands) needed for crystal metal calculations with
small unit cells. The quench is done by nonlinear conjugate gradient method.
The system is considered quenched when the energy difference between
subsequent configurations is 10${}^{-7}$ eV or less.  The DFT structure is at
$N = 150$, a number large enough to get potential energy parameters accurate
to a few percent, but small enough that convergence properties of the
calculations can be studied.

\begin{figure}
\includegraphics[width=0.9\linewidth]{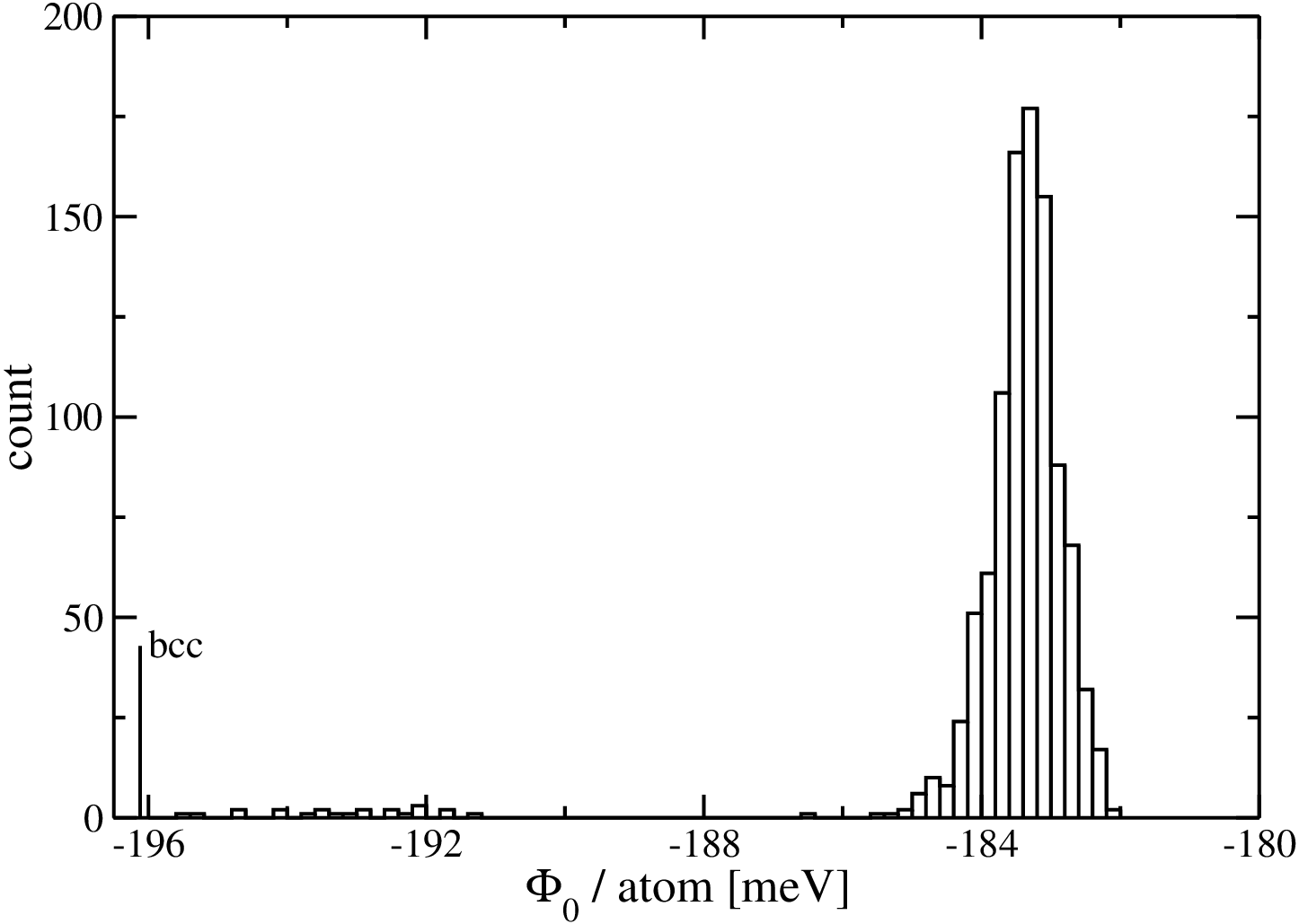}
\caption{
  \label{fig:distributionStochastic}
  Potential energy distribution of $N = 500$ Na atom systems after quenching
  from 1000 stochastic configurations using a nonlinear conjugate gradient
  method.}
\end{figure}

Because of the strong dominance of random valleys in the potential energy
surface, the DFT structure is expected to be random. To eliminate different
zeros of energy, we evaluate the energy difference
\begin{equation}
  \label{eq:energy_difference}
  \Delta \Phi_{0} = \Phi_{0}^{r} - \Phi_{0}^{bcc},
\end{equation}
where the superscripts $r$ and $bcc$ represent respectively the random
structure and the bcc crystal.  The comparison is
\begin{equation}
  \label{eq:mean_difference_PP_DFT}
  \Delta \Phi_{0}/N = \left\{
    \begin{array}{l}
      12.75 \mbox{ meV/atom (pair potential)} \\
      12.76 \mbox{ meV/atom (DFT)}.
    \end{array} \right.
\end{equation}
The value for the pair potential is from the second mean in
Eq.~(\ref{eq:mean_energy}), which is also calculated from stochastic
configurations. The difference of $0.01$ meV/atom is small compared to
theoretical errors in $\Delta \Phi_{0}/N$, and also compared to experimental
error in the energy of liquid Na at melt.

An independent confirmation is furnished by $\theta_0$. The Na pair potential
value is from \cite{lorenzi-venneri:041203}:
\begin{equation}
  \label{eq:theta0_PP_DFT}
  \theta_{0} = \left\{
    \begin{array}{l}
      98.4 \mbox{ K (pair potential)} \\
      97.6 \mbox{ K (DFT)}.
    \end{array} \right.
\end{equation}
The relative difference of $0.8\%$ in $\theta_0$ will make a
corresponding difference of $0.3\%$ in the theoretical entropy of
liquid Na at melt.  The difference is well within theoretical error in
$\theta_0$, and is close to the experimental error in the entropy.

The structural pair distribution $G(r)$ is not a parameter of the V-T
Hamiltonian, but $G(r)$ has a role in density fluctuation phenomena,
and it is therefore interesting to compare the DFT and pair potential
results.  The comparison is shown in Fig.~\ref{fig:gOfR}, where the agreement
is excellent. Notice the DFT curve ($N = 150$) has a small deficiency at the
tip of the first peak, compared to the pair potential curve ($N = 500$).  This
deficiency is a small--$N$ effect, and is observed also with the pair
potential at $N = 168$, but not at $N \geq 500$ (\cite{PhysRevE.59.2955},
Fig.~2).

\section{Extracting densities of states from quench results}
\label{sec:extracting}

\subsection{Quenches from equilibrium configurations}
\label{sec:quenches_from_MD}

In classical statistical mechanics, the partition function for a single
potential valley harmonically extended to infinity is $e^{-\beta \Phi_{0}}
\left( T/\theta_{0} \right)^{3N}$. The factor $\left( T/\theta_{0}
\right)^{3N}$ expresses the vibrational motion. The transit contribution, which
accounts for the valley--valley intersections, will be neglected here.  The
total liquid partition function $Z$ is
\begin{equation}
  \label{eq:partition_function}
  Z = \int \int G (\Phi_{0}, \theta_{0}) e^{-\beta \Phi_{0}} \left(
  \frac{T}{\theta_{0}} \right)^{3N} d\theta_{0} \,\, d\Phi_{0},
\end{equation}
where $G (\Phi_{0}, \theta_{0})$ is the joint density of states for the
collection of valleys. The normalization of $G (\Phi_{0}, \theta_{0})$ is
$\mathcal{N}$, the total number of valleys. The equilibrium statistical weight
of a single valley is
\begin{equation}
  \label{eq:statistical_weight}
  W_{eq} (\Phi_{0}, \theta_{0}) = \frac{e^{-\beta \Phi_{0}} \left( T/\theta_{0}
  \right)^{3N}}{Z}.
\end{equation}
In equilibrium at $T \geq T_{m}$, the probability of finding the system in
$d\theta_{0}$ at $\theta_{0}$, and in $d\Phi_{0}$ at $\Phi_{0}$, is $P
(\Phi_{0}, \theta_{0}) \,\, d\theta_{0} \,\, d\Phi_{0}$, where
\begin{equation}
  \label{eq:probability}
  P (\Phi_{0}, \theta_{0}) = G (\Phi_{0}, \theta_{0}) W_{eq} (\Phi_{0},
  \theta_{0}).
\end{equation}
Upon quenching from an equilibrium trajectory at $T \geq T_{m}$, the
structures sampled will exhibit a distribution proportional to $P (\Phi_{0},
\theta_{0})$ \cite{footnote:equilibrium}.  In view of
Eqs.~(\ref{eq:partition_function}) and (\ref{eq:statistical_weight}) it
follows that $P (\Phi_{0}, \theta_{0})$ is insensitive to the normalization of
$G(\Phi_{0}, \theta_{0})$. Therefore measurements of $P(\Phi_{0}, \theta_{0})$
can not be used to count the valleys.

\begin{figure}
\includegraphics[width=0.9\linewidth]{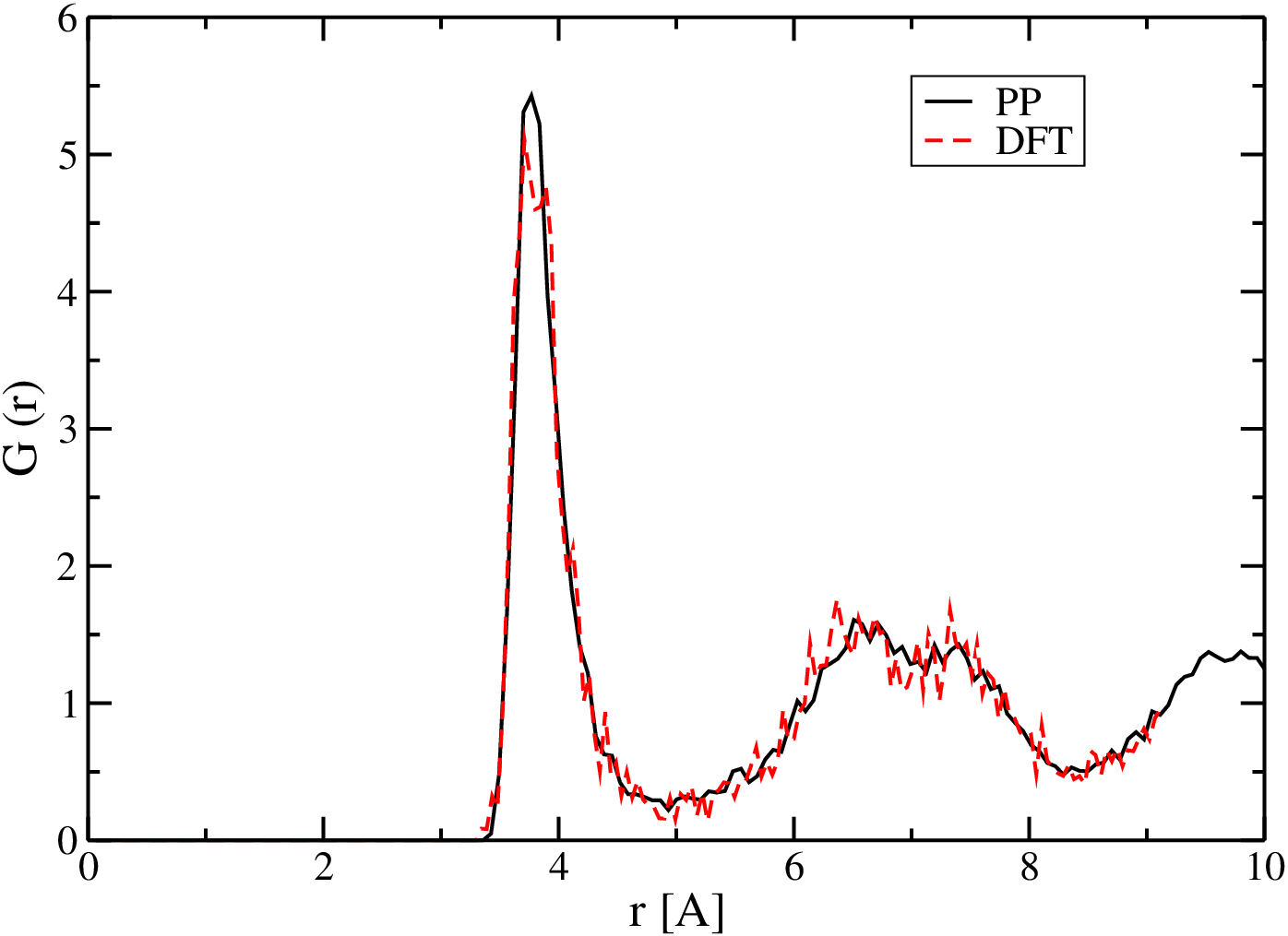}
\caption{
  \label{fig:gOfR}
  (Color online) Structural pair distribution function $G(r)$ for quenched 
  structures from pair potential (solid curve, black, $N = 500$) and a single
  DFT calculation (broken curve, red, $N = 150$).}
\end{figure}

The probability of finding the system in $d\Phi_{0}$ at $\Phi_{0}$ is $P
(\Phi_{0}) \,\, d\Phi_{0}$, where
\begin{equation}
  \label{eq:probability_definition}
  P(\Phi_{0}) = \int P(\Phi_{0}, \theta_{0}) \,\, d\theta_{0}.
\end{equation}
Upon quenching from an equilibrium trajectory at $T \geq T_{m}$, the
structures sampled will exhibit a distribution of $\Phi_{0}$ proportional to
$P (\Phi_{0})$. The distribution in Fig.~\ref{fig:distributionMD} is
proportional to $P (\Phi_{0})$.  However, the density of states in $\Phi_{0}$
is $G (\Phi_{0})$, given by
\begin{equation}
  \label{eq:density_of_states}
  G (\Phi_{0}) = \int G (\Phi_{0}, \theta_{0}) \,\, d\theta_{0},
\end{equation}
and differs from $P (\Phi_{0})$ by the statistical weight $W_{eq} (\Phi_{0},
\theta_{0})$. In our studies of liquid dynamics, the purpose of quenching is
to determine the densities of states $G (\Phi_{0})$ and $G (\Phi_{0},
\theta_{0})$, because these are parameters of the Hamiltonian. These densities
of states must be solved for from the observed distributions $P (\Phi_{0})$
and $P (\Phi_{0}, \theta_{0})$ from the above equations. Even though the
symmetric structures are supposed to be unimportant for the liquid as $N
\rightarrow \infty$, those structures will always be statistically important
at $T < T_{m}$, and will therefore be included in our analysis.

Let us introduce subscripts $r$ and $s$ for random and symmetric, respectively,
and write
\begin{equation}
  G (\Phi_{0}, \theta_{0}) = G_{r}(\Phi_{0}, \theta_{0}) + G_{s}(\Phi_{0},
  \theta_{0}),
\end{equation}
and correspondingly $Z = Z_{r} + Z_{s}$. From Figs.~\ref{fig:distributionMD}
and \ref{fig:distributionStochastic}, $P_{r}(\Phi_{0})$ at $N = 500$ has very
small width, comparable to experimental error in the internal energy of the
liquid at melt, and this width is expected to decrease further as $N$
increases \cite{N-Dependent-Quenches}. These results suggest a model for
$G_{r} (\Phi_{0}, \theta_{0})$. Let us define the liquid $\Phi_{0}^{l}$ as the
mean $\Phi_{0}$ for random structures when $N \rightarrow \infty$, with a
similar definition for $\theta_{0}^{l}$. The model is
\begin{eqnarray}
  G_{r} (\Phi_{0}, \theta_{0}) & = & \mathcal{N}_{r} \,\, \delta \left(
  \Phi_{0} - \Phi_{0}^{l} \right) \delta \left( \theta_{0} - \theta_{0}^{l}
  \right) \left[ 1 + \mathcal{O} \left( N^{-\alpha} \right) \right], \nonumber \\
  \label{eq:model}
  G_{r} (\Phi_{0}) & = & \mathcal{N}_{r} \,\, \delta \left( \Phi_{0} -
  \Phi_{0}^{l} \right) \left[ 1 + \mathcal{O} \left( N^{-\alpha} \right)
  \right],
\end{eqnarray}
where $\alpha > 0$, and $\mathcal{N}_{r}$ is the number of random valleys.
From this it follows that $Z_{r} = \mathcal{N}_{r} \,\, e^{-\beta
\Phi_{0}^{l}} \left( T / \theta_{0}^{l} \right)^{3N}$, and the random
contributions to Eqs.~(\ref{eq:probability}) and
(\ref{eq:probability_definition}) become
\begin{eqnarray}
  P_{r} (\Phi_{0}, \theta_{0}) & = & \frac{\delta \left( \Phi_{0} - \Phi_{0}^{l}
  \right) \delta \left( \theta_{0} - \theta_{0}^{l} \right)}{1 + Z_{s}/Z_{r}}
  \left[ 1 + \mathcal{O} \left( N^{-\alpha} \right) \right], \nonumber \\
  \label{eq:random_probability}
  P_{r} (\Phi_{0}) & = & \frac{\delta \left( \Phi_{0} - \Phi_{0}^{l} \right)}{1
  + Z_{s}/Z_{r}} \left[ 1 + \mathcal{O} \left( N^{-\alpha} \right) \right].
\end{eqnarray}
Hence, to finite--$N$ errors, the random valley Hamiltonian parameters
$\Phi_{0}^{l}$ and $\theta_{0}^{l}$ are determined directly by quenches from
equilibrium MD at $T \geq T_{m}$. And, because of the form of
Eq.~(\ref{eq:model}) for $G_{r} (\Phi_{0}, \theta_{0})$, these observations
will remain true when the statistical mechanics theory is improved to include
transit effects.

The symmetric density $G_{s} (\Phi_{0}, \theta_{0})$ apparently has
$N$--independent width with $\Phi_{0}$ ranging from $\Phi_{0}^{bcc}$ to
$\Phi_{0}^{l}$. Symmetric structures with $\Phi_{0} > \Phi_{0}^{l}$ exist, but
they are rare for monatomic systems. The $\theta_{0}$ dependence of $G_{s}
(\Phi_{0}, \theta_{0})$ is not trivial. One expects that additional (symmetry)
parameters are important for symmetric structures.  Nevertheless, $G_{s}
(\Phi_{0})$ and $G_{s} (\Phi_{0}, \theta_{0})$ are well defined, and can be
extracted from quench data with the aid of Eqs.~(\ref{eq:probability}) and
(\ref{eq:probability_definition}).

\subsection{Quenches from stochastic configurations}
\label{sec:extracting_stochastic_quenches}

On an equilibrium trajectory at $T \geq T_{m}$, the probability the system is
found in a given potential energy valley is $W_{eq} (\Phi_{0}, \theta_{0})$ for
that valley. The statistical weight is quite different for stochastic
configurations. These configurations are uniformly distributed over
configuration space, except for the small excluded Cartesian--space volume at
each nucleus. Neglecting this constraint, the probability the system is found in
a given potential valley is the valley volume divided by the entire
$3N$--dimensional volume. Let us denote the corresponding statistical weight
factors as $W_{r} (\Phi_{0}, \theta_{0})$ and $W_{s} (\Phi_{0}, \theta_{0})$
for random and symmetric valleys respectively.

Because of their uniformity, the random valleys all have the same volume in
the thermodynamic limit. To arrive at a complete solution, it is necessary to
include the symmetric valleys. Let us assume that they also have a uniform
configuration space volume. The number of random (symmetric) valleys is
denoted $\mathcal{N}_{r}$ ($\mathcal{N}_{s}$), and the
single--valley volume is $\mathcal{V}_{r}$ ($\mathcal{V}_{s}$).  The
statistical single--valley weights are
\begin{eqnarray}
  W_{r} & = & \frac{\mathcal{V}_{r}}{\mathcal{N}_{r} \mathcal{V}_{r} +
  \mathcal{N}_{s} \mathcal{V}_{s}}, \\
  W_{s} & = & \frac{\mathcal{V}_{s}}{\mathcal{N}_{r} \mathcal{V}_{r} +
  \mathcal{N}_{s} \mathcal{V}_{s}}.
\end{eqnarray}
The probability distributions are
\begin{eqnarray}
  P_{r} (\Phi_{0}, \theta_{0}) & = & G_{r} (\Phi_{0}, \theta_{0}) W_{r}, \\
  P_{s} (\Phi_{0}, \theta_{0}) & = & G_{s} (\Phi_{0}, \theta_{0}) W_{s}.
\end{eqnarray}
Hence the random and symmetric densities of states are each proportional to the
distribution found in quenches from stochastic configurations. Applying the
model of Eq.~(\ref{eq:model}) for $G_{r} (\Phi_{0}, \theta_{0})$ yields
\begin{equation}
  P_{r} (\Phi_{0}, \theta_{0}) = \frac{\delta \left(\Phi_{0} - \Phi_{0}^{l}
  \right) \delta \left( \theta_{0} - \theta_{0}^{l} \right)}{1 + \left(
  \mathcal{N}_{s} \mathcal{V}_{s} / \mathcal{N}_{r} \mathcal{V}_{r} \right)}
  \left[ 1 + \mathcal{O} \left( N^{-\alpha} \right) \right].
\end{equation}
The conclusion here is the same as with equilibrium configurations,
Eq.~(\ref{eq:random_probability}), that the parameters $\Phi_{0}^{l}$ and
$\theta_{0}^{l}$ are determined directly by quenches from stochastic
configurations, up to finite--$N$ errors. The reason, of course, is the form
of the model for $G_{r} (\Phi_{0}, \theta_{0})$, Eq.~(\ref{eq:model}). For
symmetric structures, the above equations reveal two significant points:
\begin{enumerate}
  \item
    Quenches from stochastic configurations can determine the magnitude of
    $G_{s} (\Phi_{0}, \theta_{0})$ relative to $G_{r} (\Phi_{0}, \theta_{0})$,
    but only when $W_{s}/W_{r}$ is known.
  \item
    The relation between $G_{r} (\Phi_{0}, \theta_{0})$ as determined by the
    two quench methods is unknown until $W_{s}$ is evaluated.
\end{enumerate}
These points are relevant to the distributions shown in
Figs.~\ref{fig:distributionMD} and \ref{fig:distributionStochastic}.

\section{Conclusions}
\label{sec:conclusions}

In this article, we have addressed two main research goals: (1) The use of
stochastic configurations to probe the distribution of potential energy
minima, and (2) the calibration of the V-T Hamiltonian with DFT.

\subsection{Stochastic Configurations}

Quenches from stochastic configurations produce statistically
indistinguishable distributions of the potential energy, $\Phi_{0}/N$,
compared to quenches from equilibrium liquid MD trajectories.  We have
demonstrated that quenching from stochastic configurations can be used to find
the entire distribution of Na potential energy minima, i.e. they reliably
reproduce the correct distribution of random and symmetric structures. This is
illustrated by a comparison of $\Phi_{0}/N$ distributions for quenches from
equilibrium MD and from stochastic configurations
(Figs.~\ref{fig:distributionMD} and \ref{fig:distributionStochastic},
Sec.~\ref{sec:validatestochastic}). Quenches from stochastic configurations
yield an accurate random distribution for Na in agreement with the random
distribution from quenches from equilibrium MD, Eq.~(\ref{eq:mean_energy}).

Compared to generating and selecting configurations from equilibrium MD
trajectories, our stochastic configuration method is significantly faster and
more economical (Sec.~\ref{sec:generating}).  Stochastic configurations do not
require interatomic potentials or costly equilibration and long simulation
times. Simply generate random Cartesian coordinates for each atom under a
minimal excluded--volume constraint to eliminate particle overlap.  Hence,
this procedure requires very little computational effort, an economy that
accommodates \emph{ab initio} quench methods even for large systems.

\subsection{Calibration of the V-T Hamiltonian}

Calibration of the V-T Hamiltonian is based on the presumed dominance and
uniformity of random valleys as $N \rightarrow \infty$. This view is given
mathematical expression in the model for $G_{r} (\Phi_{0}, \theta_{0})$,
Eq.~(\ref{eq:model}). It follows that the thermodynamic limit parameters
$\Phi_{0}^{l}$ and $\theta_{0}^{l}$ are determined directly from data for
either MD quenches or stochastic quenches, up to finite--$N$ errors.

We have demonstrated that the DFT structure in Sec.~\ref{sec:calibrateH} is
random by comparing the mean potential energy $\Phi_{0}/N$ with the pair
potential results in Eq.~(\ref{eq:mean_difference_PP_DFT}), the phonon moment
$\theta_{0}$ in Eq.~(\ref{eq:theta0_PP_DFT}), and the pair distribution
function $G(r)$ in Fig.~\ref{fig:gOfR}. We conclude that being random, the
DFT structure can provide \emph{ab initio} calibration of the V-T
Hamiltonian.

As verified by their pair distribution function, stochastic configurations
have nuclei distributed nearly uniformly over the system volume
(Figs.~\ref{fig:gOfR_constant_binsize} and \ref{fig:gOfR_constant_volume},
Sec.~\ref{sec:generating}). Therefore among stochastic configurations, the
statistical weight for a many--atom potential energy valley is (nearly)
proportional to the valley volume. This is in contrast to quenches from
equilibrium MD, which require extensive modeling to extract the Boltzmann
factor from the weight \cite{0953-8984-12-29-325}. Given the statistical
weight, characteristics of the underlying potential surface can be extracted
from data acquired by stochastic quenches (Sec.~\ref{sec:extracting}).

We do not suggest that DFT quenches from stochastic configurations will
invariably arrive at random structures, just as quenches from an equilibrium
MD trajectory may result in a symmetric structure. Indeed, some symmetric
structures have appeared in our DFT quenches.  Precisely what is required to
eliminate symmetric structures from any collection of quench data is an
ongoing research question.

\section{Acknowledgments}

EH and RL would like to thank Fondecyt projects 11070115 and 11080259. They
are also thankful for the support of ``Proyecto Anillo'' ACT-24.  This work was
funded by the US Department of Energy under contract number DE-AC52-06NA25396.
The open and friendly scientific atmosphere of the Ten Bar International
Science Caf\'{e} is also acknowledged.

\bibliography{article}

\begin{thebibliography}{31}
\bibitem[{foo()}]{footnote:equilibrium}
\bibinfo{note}{An equilibrium MD system moves at constant total energy and
  total momentum. These constraints affect fluctuations in order 1, but affect
  mean values only in order $N^{-1}$ \cite{PhysRev.153.250}. We exercise
  caution when using canonical weighting to analyze MD results.}
\end{thebibliography}

\end{document}